\documentclass[showpacs,twocolumn,superscriptaddress]{revtex4}

\usepackage{amsmath,amssymb,mathrsfs}
\usepackage[dvips]{graphicx}
\usepackage{psfrag}
\usepackage{color}

\newcommand{\up}{\uparrow}
\newcommand{\dw}{\downarrow}
\newcommand{\ket}[1]{\left|#1\right\rangle}
\newcommand{\bra}[1]{\left\langle#1\right|}
\newcommand{\sgn}[1]{\text{sgn}(#1)\,}
\newcommand{\ii}{\text{i}\,}

\def\ea{\emph{et al.}}

\def\prl#1#2#3{Phys.\ Rev.\ Lett.\ {\bf #1}, #2 (#3)}

\def\prb#1#2#3{Phys.\ Rev.\ B {\bf #1}, #2 (#3)}

\def\science#1#2#3{Science {\bf #1}, #2 (#3)}

\def\C60{A$_x$C$_{60}$}

\def\HgCu3{HgCa$_2$Cu$_3$O$_{8+y}$}
\def\HgCu4{HgBa$_2$Ca$_3$Cu$_4$O$_{10+y}$}
\def\TlCu{Tl$_2$Ba$_2$CuO$_{6+\delta}$}
\def\TlCu3{Tl$_2$Ba$_2$Ca$_2$Cu$_3$O$_{10+y}$}
\def\TlCu4{Tl$_2$Ba$_2$Ca$_3$Cu$_4$O$_{12+y}$}

\def\BiCu3{Bi$_2$Sr$_2$Ca$_{2}$Cu$_3$O$_y$}

\def\8LSCO{La$_{1.88}$Sr$_{.12}$CuO$_4$}
\def\110LNSCO{La$_{1.5}$Nd$_{0.4}$Sr$_{0.1}$CuO$_{4}$}
\def\stage4LCO{La$_{2}$CuO$_{4+\delta}$}
\def\Y248{YBa$_2$Cu$_4$O$_8$}

\def\hts{high temperature superconductors}

\def\NbSe2{NbSe$_2$}
\def\TaSe2{TaSe$_2$}
\def\TiSe2{TiSe$_2$}
\def\NaCoOH2O{Na$_{0.3}$CoO$_{2y}$H$_2$O}
\def\MgB2{MgB${}_2$}

\allowdisplaybreaks[4]

\begin{document}
%%%%%%%%%%%%%%%%%%%%%%%%%%%%%%%%%%%%%%%%%%%%%%%%%%%%%%%%%%%%%%%%%%%%%%%
\title{Local density of states of 1D Mott insulators and
CDW states with a boundary}  
%%%%%%%%%%%%%%%%%%%%%%%%%%%%%%%%%%%%%%%%%%%%%%%%%%%%%%%%%%%%%%%%%%%%%%%
\author{Dirk Schuricht}
\affiliation{The Rudolf Peierls Centre for Theoretical Physics, 
University of Oxford, 1 Keble Road, OX1 3NP, Oxford, UK}
\author{Fabian H. L. Essler}
\affiliation{The Rudolf Peierls Centre for Theoretical Physics, 
University of Oxford, 1 Keble Road, OX1 3NP, Oxford, UK}
\author{Akbar Jaefari}
\affiliation{Department of Physics, University of Illinois at Urbana-Champaign,
1110 W. Green St, Urbana, IL 61801-3080, USA}
\author{Eduardo Fradkin}
\affiliation{Department of Physics, University of Illinois at Urbana-Champaign,
1110 W. Green St, Urbana, IL 61801-3080, USA}
\date{\today}

\pagestyle{plain}
\begin{abstract}
We determine the local density of states (LDOS) of one-dimensional
incommensurate charge density wave (CDW) states in the presence of a strong
impurity potential, which is modeled by a boundary. We find that the CDW
gets pinned at the impurity, which results in a singularity in the Fourier
transform of the LDOS at momentum $2k_\mathrm{F}$. At energies above the
spin gap we observe dispersing features associated with the spin and charge
degrees of freedom respectively. In the presence of an impurity magnetic
field we observe the formation of a bound state localized at the impurity.
All of our results carry over to the case of 1D Mott insulators
by exchanging the roles of spin and charge degrees of freedom. We discuss
the implications of our result for scanning tunneling microscopy experiments
on spin-gap systems such as two-leg ladder cuprates.
\end{abstract}
%68.37.Ef STM
%71.10.Pm fermions in low D
%72.80.Sk Insulators
\maketitle
%%%%%%%%%%%%%%%%%%%%%%%%%%%%%%%%%%%%%%%%%%%%%%%%%%%%%%%%%%%%%%%%%%%%%%%

In recent years scanning tunneling microscopy (STM) and spectroscopy (STS)
techniques have proved to be very useful tools for studying strongly
correlated electron systems such as {\hts} (HTSC)~\cite{STMhightc}, carbon
nanotubes~\cite{Odom-02}, and rare-earth compounds~\cite{STMrare}. The usage
of magnetic tips has also enabled the investigation of magnetic
properties~\cite{magnSTM}. These experiments have motivated an intense
research effort on the theory of STS~\cite{wang-lee,Kivelson-03}, the main
focus being on quasiparticle properties in HTSC. STM measures the local
differential conductance $dI/dV(x)$, which is directly related to the LDOS.
Impurities break translational invariance and lead to a modification of the
LDOS in their vicinity, from which one can infer characteristic properties of
the bulk state of matter as well as the nature of its electronic excitations.
Interestingly this holds even in strongly correlated systems without
quasiparticle excitations as was demonstrated for a non-magnetic impurity
placed in a gapless Luttinger liquid \cite{Eggert00,Kivelson-03}, which can be
regarded as a quantum critical CDW state. In this case, at low energies the
impurity behaves effectively as a physical boundary \cite{kane-fisher} where
the phase of the CDW order parameter gets pinned, giving rise to induced CDW
order. 

Here we consider STS in a 1D strongly correlated system with a spin gap in the
presence of an impurity. This problem is of interest to the study of (quasi)
1D CDW systems, two-leg ladder materials with strong superconducting
correlations, and the stripe phases of HTSC \cite{stripe}. We focus on the
regime in which the scattering at the impurity is strong and hence at
sufficiently low energies it acts as a physical boundary~\cite{tsvelik}. The
inherently non-perturbative nature of STS in gapped (quasi)-1D systems
requires an entirely different treatment compared to previously studied cases.
Following Ref.~[\onlinecite{Kivelson-03}] we consider the spatial Fourier
transform of the LDOS throughout, as this allows physical properties to be
more easily identified and is commonly used in
experiments~\cite{STMhightc,STMrare}.

The starting point of our analysis is the continuum description of a CDW
state. The latter arises in two rather different kinds of 1D correlated
electron systems: (1) A partially filled band of spinful electrons coupled to
optical phonons of frequency $\omega_{\rm ph}$. At energies small compared to
$\omega_{\rm ph}$ the electron-phonon coupling results in an attractive
interaction between electrons that can overcome the Coulomb repulsion
~\cite{HirschFradkin83}.  (2) Strongly correlated two- and three-leg ladder
systems~\cite{white-noack-scalapino,stripe}. Here, in spite of strongly
repulsive electron interactions, a Mott state with a finite spin gap occurs
for a range of dopings around half filling~\cite{endnote}. In both cases 
%(1) and (2) 
there is a broad range of parameters such that at low energies the system
gives rise to a CDW state characterized by a gapped spin sector and a gapless
charge sector.
Regardless of the microscopic origin of the spin gap, by taking
the continuum limit and bosonizing one arrives at a spin-charge
separated theory describing collective charge and spin degrees of
freedom. We now imagine a potential impurity to be present which 
at low energies and temperatures effectively cuts the system into two
disconnected parts \cite{endnote3}. We then can model the impurity
by a boundary condition on the continuum electron field
$\Psi_\sigma(x=0)=0$, resulting in a spin-charge separated Hamiltonian
of the form $H=H_\mathrm{c}+H_\mathrm{s}$ 
\begin{eqnarray}
H_\mathrm{c}&=&\frac{v_\mathrm{c}}{16\pi}\int_{-\infty}^0 dx 
\biggl[\frac{1}{K_\mathrm{c}^2}\bigr(\partial_x\Phi_\mathrm{c}\bigr)^2+
K_\mathrm{c}^2\bigr(\partial_x\Theta_\mathrm{c}\bigr)^2\biggr],
\label{eq:chargehamiltonian}\\
H_\mathrm{s}&=&\frac{v_\mathrm{s}}{16\pi}\int_{-\infty}^0 dx 
\biggl[\frac{1}{K_\mathrm{s}^2}\bigr(\partial_x\Phi_\mathrm{s}\bigr)^2+
K_\mathrm{s}^2\bigr(\partial_x\Theta_\mathrm{s}\bigr)^2\biggr]
\nonumber\\
& &-\frac{g_\mathrm{s}}{(2\pi)^2}
\int_{-\infty}^0 dx\, \cos\Phi_\mathrm{s}.
\label{eq:spinhamiltonian}
\end{eqnarray}
Here $\Phi_\mathrm{c,s}$ are canonical Bose fields which 
%in the simplest case
satisfy the boundary conditions $\Phi_\mathrm{c,s}(x=0)=0$, and
$\Theta_\mathrm{c,s}$ are their dual fields.  The charge and spin velocities
$v_\mathrm{c,s}$, the Luttinger parameters $K_\mathrm{c,s}$ and coupling
constant $g_\mathrm{s}$ are functions of the parameters defining the
underlying microscopic model. In the case of single chain electron-phonon
systems we have $v_\mathrm{s}>v_\mathrm{c}$ and $K_\mathrm{c}>1$ as we are
dealing with attractive electron-electron interactions. On the other hand, in
the case of two-leg ladders there is a spin gap with
$v_\mathrm{s}<v_\mathrm{c}$ and $K_\mathrm{c}<1$.  We will consider both
situations in what follows.  The charge sector (\ref{eq:chargehamiltonian})
describes gapless collective charge excitations propagating with velocity
$v_\mathrm{c}$, whereas the spin excitations are described by a
sine-Gordon model with a boundary (\ref{eq:spinhamiltonian}). In the
regime considered here 
the bulk spectrum of the latter consists of gapped (anti)soliton excitations.
At the Luther-Emery point $K_\mathrm{s}=1/\sqrt{2}$ the spin sector is
equivalent to a free massive Dirac fermion~\cite{LutherEmery74}. As is well
known, \eqref{eq:chargehamiltonian}--\eqref{eq:spinhamiltonian} reduces to the
low-energy theory of a half-filled one-band Mott insulator, provided we
interchange charge and spin sector and then set $K_\mathrm{s}=1$ and
$k_\mathrm{F}=\pi/2$. By virtue of this connection all our results for CDW
states carry over to 1D Mott insulators.

The central object of our study is the time-ordered Green's
function in Euclidean space,
\begin{equation}
\label{eq:GF}
G_{\sigma\sigma'}(\tau,x_1,x_2)=
-\bra{0_\mathrm{b}}\mathcal{T}_\tau\,\Psi_\sigma(\tau,x_1)\,
\Psi_{\sigma'}^\dagger(0,x_2)\ket{0_\mathrm{b}}.
\end{equation}
Here $\ket{0_\mathrm{b}}$ is the ground state and
$\tau=\ii t$ denotes imaginary time. At low energies the electron
annihilation operator can be decomposed into right- and left-moving
components as $\Psi_{\sigma}(x)=e^{\ii k_\mathrm{F} x} R_\sigma(x)+
e^{-\ii k_\mathrm{F} x} L_\sigma(x)$, which reduces \eqref{eq:GF} to
\begin{equation}
\label{eq:GFlowenergy}
\begin{split}
G_{\sigma\sigma'}=&\;
e^{\ii k_\mathrm{F}(x_1-x_2)}\,G^{RR}_{\sigma\sigma'}
+e^{-\ii k_\mathrm{F}(x_1-x_2)}\,G^{LL}_{\sigma\sigma'}\\
&+e^{\ii k_\mathrm{F}(x_1+x_2)}\,G^{RL}_{\sigma\sigma'}
+e^{-\ii k_\mathrm{F}(x_1+x_2)}\,G^{LR}_{\sigma\sigma'},
\end{split}
\end{equation}
where e.g. $
G^{RL}_{\sigma\sigma'}=-\bra{0_\mathrm{b}}\mathcal{T}_\tau\, 
R_\sigma(\tau,x_1)\,L^\dagger_{\sigma'}(0,x_2)\ket{0_\mathrm{b}}$.  As
we are interested in the LDOS, we ultimately want to set $x_1=x_2$.
As was noted in Ref.~[\onlinecite{Kivelson-03}] it is useful to
consider the Fourier transform of the LDOS as physical properties can
be more easily identified. In momentum space the $RL$ and $LR$
contributions occur in a different region ($Q\approx \pm
2k_\mathrm{F}$) compared to the $RR$ and $LL$ parts ($Q\approx 0$). In
absence of a boundary we have
$G^{RL}_{\sigma\sigma'}=G^{LR}_{\sigma\sigma'}=0$ as the charge parts
of these Green's functions vanish. In presence of a boundary left and
right sectors are coupled and the Fourier transform of the Green's
function (\ref{eq:GFlowenergy}) concomitantly acquires a nonzero
component at $Q\approx \pm 2k_\mathrm{F}$, which gives a particularly
clean way of investigating impurity effects. For this reason we focus
on the $2k_\mathrm{F}$-part of the Green's function in what follows
but note that the small momentum regime can be analyzed analogously.

The Green's function $G_{\sigma\sigma'}^{RL}$ factorizes into a product of
correlation functions in the spin and charge sectors. The charge part can be
determined by a standard mode expansion \cite{FabrizioGogolin95}.  The
sine-Gordon model on the half-line \eqref{eq:spinhamiltonian} is known
to be integrable for quite general boundary
conditions~\cite{GhoshalZamolodchikov94}.  This enables 
us to calculate the correlation functions in the spin sector using the
boundary state formalism introduced by Ghoshal and
Zamolodchikov~\cite{GhoshalZamolodchikov94} together with a form-factor
expansion~\cite{LukyanovZamolodchikov01}. Taking into account only the leading
terms we arrive at ($\tau>0$)
\begin{equation}
G^{RL}_{\sigma\sigma'}(\tau,x,x)=
g_\mathrm{c}(\tau,x)\,g_\mathrm{s}(\tau,x),
\end{equation}
\begin{eqnarray}
g_\mathrm{c}(\tau,x)&=&-\frac{\delta_{\sigma\sigma'}}{2\pi}
\frac{1}{\bigl(v_\mathrm{c}\tau-2\ii x\bigr)^a}\,
\frac{1}{\bigl(v_\mathrm{c}\tau+2\ii x\bigr)^b}\,
\Biggl[\frac{4x^2}{v_\mathrm{c}^2\tau^2}\Biggr]^c,\nonumber\\
g_\mathrm{s}(\tau,x)&=&Z_1\,e^{\ii\frac{\pi}{4}}\Biggl[
\frac{1}{\pi}\,K_0\bigl(\Delta\tau\bigr)
+\int_{-\infty}^{\infty}\frac{d\theta}{2\pi}\,
K\bigl(\theta+\ii\tfrac{\pi}{2}\bigr)\nonumber\\
&&\quad\times\ \,e^{\theta/2}\,
e^{2\ii\tfrac{\Delta}{v_\mathrm{s}}x\sinh\theta}\,
e^{-\Delta\tau\cosh\theta}+\ldots\Biggr],
\label{eq:GRLspin}
\end{eqnarray}
where $g_\mathrm{c,s}(\tau,x)$ are the contributions of the charge and spin
sectors respectively.  Here $K_0$ is a modified Bessel function, the
normalization constant $Z_1$ was obtained in
Ref.~\cite{LukyanovZamolodchikov01}, and explicit expressions for the boundary
reflection amplitude $K(\theta)$ are given in
Ref.~\cite{GhoshalZamolodchikov94} (at the Luther-Emery point we
have~\cite{Ameduri-95} $K(\theta)=\ii\tanh\frac{\theta}{2}$). The exponents in
the charge sector are related to the Luttinger parameters by
$a=(K_\mathrm{c}+1/K_\mathrm{c})^2/8$, $b=(K_\mathrm{c}-1/K_\mathrm{c})^2/8$,
and $c=(1/K_\mathrm{c}^2-K_\mathrm{c}^2)/8$.  The subleading terms in
\eqref{eq:GRLspin} involve three or more particles in the intermediate state
or higher orders in the boundary $K$-matrix~\cite{endnote2}.
The Fourier transform of the LDOS for $E>0$ is 
\begin{equation}
N_\sigma(E,Q)=-\int_{-\infty}^0dx\int^\infty_{-\infty}\frac{dt}{2\pi}
e^{\ii(E t-Qx)}\,G_{\sigma\sigma}(t,x,x),
\label{eq:defNp}
\end{equation}
where the Green's function has been analytically continued to real time.
For $Q\approx 2k_\mathrm{F}$ only $G^{RL}_{\sigma\sigma}$ contributes
and using \eqref{eq:GRLspin} we arrive at our main result
\begin{eqnarray}
&&N_\sigma(E,2k_\mathrm{F}+q)\propto
-\Theta(E-\Delta)\sum_{i=1}^2 N_i(E,q),\nonumber\\
&&N_i(E,q)=\int_{-A}^A d\theta
\frac{h_i(\theta)\,u_i^{2c+1}}{(E\!-\!\Delta\cosh\theta)^{2-a-b}}
\nonumber\\
&&\hskip 1.5cm\times F_1\bigl(2c+1,a,b,a+b+2c;u_i^*,-u_i\bigr).
\label{eq:Ni}
\end{eqnarray}
Here $|q|\ll 2k_\mathrm{F}$, $A=\mathrm{arcosh}\bigl(\tfrac{E}{\Delta}\bigr)$,
$F_1$ denotes Appell's hypergeometric function,
$h_1(\theta)=1$,
$h_2(\theta)=K\bigl(\theta+\ii\tfrac{\pi}{2}\bigr)\,e^{\theta/2}$,
$u_1=2(E-\Delta\cosh\theta)/v_\mathrm{c}q+\ii\sgn{v_\mathrm{s}q/\Delta}\delta$,
and $u_2=2v_\mathrm{s}(E-\Delta\cosh\theta)/
v_\mathrm{c}(v_\mathrm{s}q-2\Delta\sinh\theta)+
\ii\sgn{v_\mathrm{s}q/\Delta-2\sinh\theta}\delta$, where $\delta\rightarrow
0+$. 
%The result \eqref{eq:Ni} is valid for $a+b<2$ and $-1/2<c$. 
Below we plot %$N_\sigma(E,2k_\mathrm{F}+q)$ 
\eqref{eq:Ni} for two different parameter regimes. We smoothen the
singularities in the LDOS by taking $\delta$ small but finite. In experiments
the singularities are broadened by instrumental resolution and temperature.
The results presented below apply to the regime $T\ll
E,\Delta,v_\mathrm{c}/a_0$ ($a_0$ is the lattice spacing), where temperature
effects are negligible.
\begin{figure}[t]
\centering
\includegraphics[scale=0.34,clip=true]{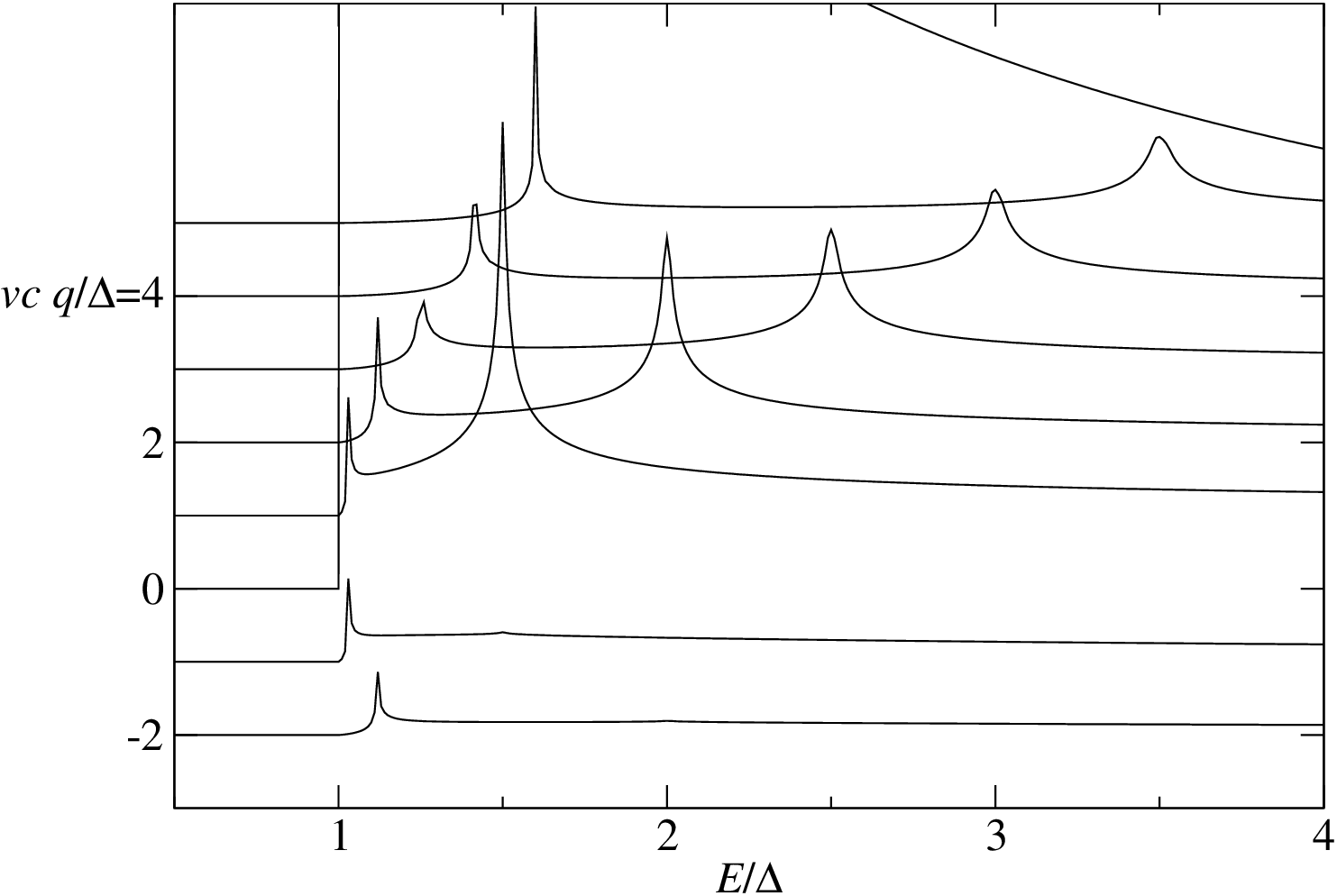}
\caption{$|N_\sigma(E,2k_\mathrm{F}+q)|$ (arbitrary units) for 
  $K_\mathrm{c}=0.8$, $K_\mathrm{s}=1$ and $v_\mathrm{c}=2v_\mathrm{s}$. The
  curves are constant $q$-scans which have been offset along the y-axis by a
  constant with respect to one another.  We observe dispersing features at
  $E_\mathrm{s}=\sqrt{\Delta^2+(v_\mathrm{s}q/2)^2}$ and
  $E_\mathrm{c}=v_\mathrm{c}|q|/2+\Delta$.  For $q<0$ the dispersing features
  are strongly suppressed.}
\label{fig:plot1}
\end{figure}

{\it Repulsive Case:} We first consider the case of repulsive electron
interactions ($v_\mathrm{s}<v_\mathrm{c}$, $K_\mathrm{c}<1$). In
Fig.~\ref{fig:plot1} we plot $N_\sigma(E,2k_\mathrm{F}+q)$ for the case of
unbroken spin rotational symmetry ($K_\mathrm{s}=1$). The LDOS is dominated by
the strong peak at momentum $2k_\mathrm{F}$, which has its origin in
$N_1(E,Q)$ and is indicative of the CDW order being pinned at the boundary.
This is analogous to the Luttinger liquid case \cite{Kivelson-03}.  At low
energies above the spin gap $\Delta$ we further observe two dispersing
features associated with the collective spin and charge degrees of freedom
respectively: (1) a ``charge peak'' that follows
$E_\mathrm{c}=v_\mathrm{c}|q|/2+\Delta$ and (2) a ``spin peak'' at position
$E_\mathrm{s}=\sqrt{\Delta^2+(v_\mathrm{s}q/2)^2}$.  The charge peak arises
from the contribution $N_1(E,Q)$ to the Fourier transform of the LDOS,
whereas the spin peak has its origin in $N_2(E,Q)$, which encodes the
effects of the boundary on the spin degrees of freedom.  We note that like in
the Luttinger liquid case~\cite{Kivelson-03} the phase of $N_\sigma$
exhibits characteristic jumps at the peak positions.

\begin{figure}[t]
\centering
\includegraphics[scale=0.34,clip=true]{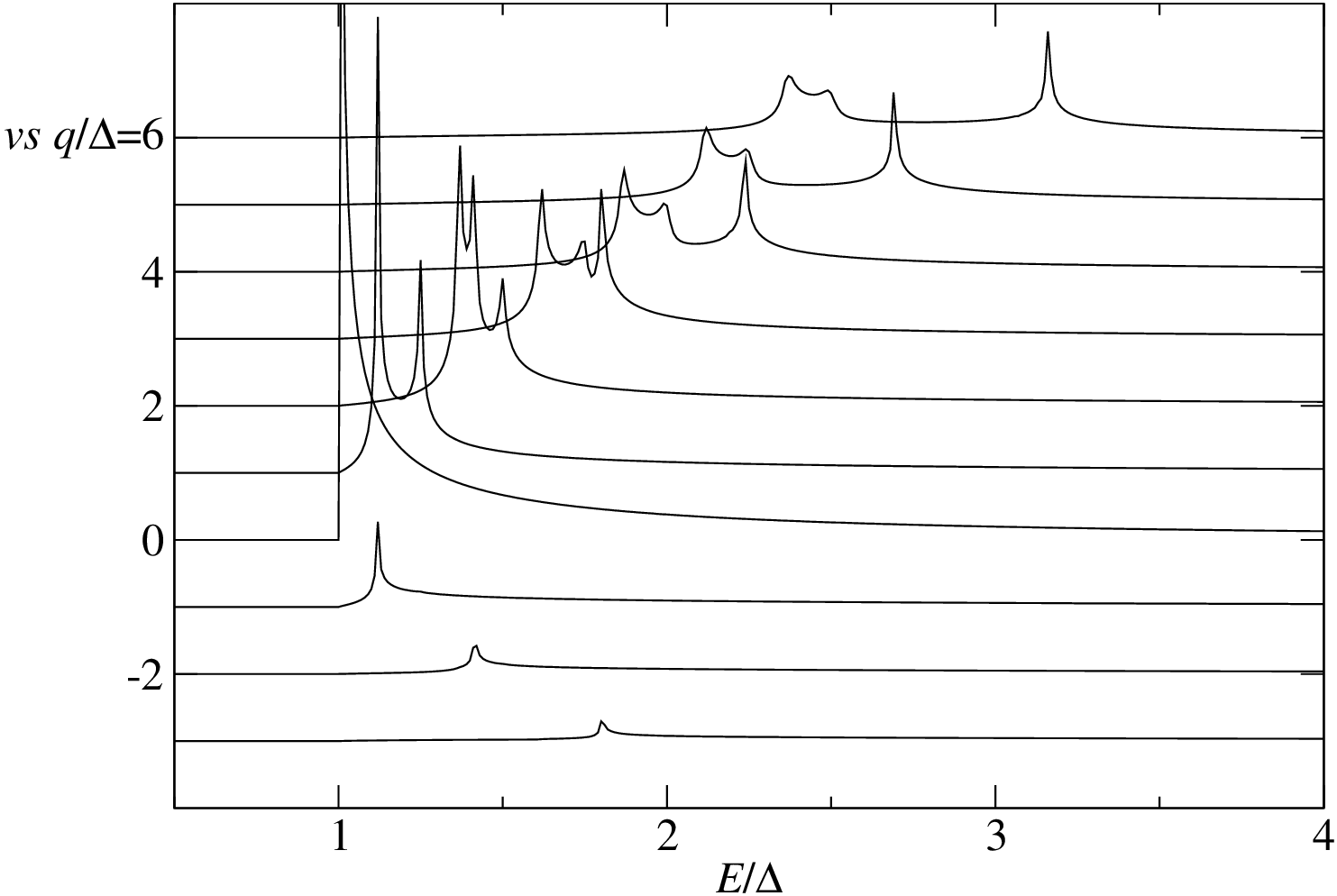}
\caption{$|N_\sigma(E,2k_\mathrm{F}+q)|$ (arbitrary units) for 
  $K_\mathrm{c}=1.2$, $K_\mathrm{s}=1$ and $v_\mathrm{s}=2v_\mathrm{c}$.  We
  observe dispersing features at $E_\mathrm{s}$, $E_\mathrm{c}$, and
  $E=v_\mathrm{c}|q|/2+\Delta\sqrt{1-(v_\mathrm{c}/v_\mathrm{s})^2}$ (for
  $|q|>q_0$ only). }
\label{fig:plot2}
\end{figure}

{\it Attractive Case:} In Fig.~\ref{fig:plot2} we show
$N_\sigma(E,2k_\mathrm{F}+q)$ for the case of attractive electron interactions
($v_\mathrm{s}>v_\mathrm{c}$, $K_\mathrm{c}>1$) and unbroken spin rotational
symmetry. The peak at $2k_\mathrm{F}$ is much less pronounced than in the
repulsive case.  We again observe charge and spin peaks that follow
$E_\mathrm{c}$ and $E_\mathrm{s}$ respectively.  For momenta $q$ above a
critical value $q_0=2\Delta
v_\mathrm{c}/v_\mathrm{s}\sqrt{v_\mathrm{s}^2-v_\mathrm{c}^2}$ a third
dispersing low-energy peak appears at
$E=v_\mathrm{c}|q|/2+\Delta\sqrt{1-(v_\mathrm{c}/v_\mathrm{s})^2}$.  This
feature can be thought of as arising from a spin excitation with momentum
$q_0$ and a charge excitation with momentum $q-q_0$. This is reminiscent of
what is found for the single-particle spectral function in the bulk
\cite{ET03}.

So far we have considered the simplest possible boundary conditions
corresponding to a spin-independent phase shift $\pi$. In reality one may
expect an impurity to give rise to a local potential or local magnetic field,
which result in more general phase shifts upon reflection of particles at the
boundary. In particular, these more general boundary conditions can give rise
to boundary bound states, see e.g.~\cite{BBS}. In order to exhibit the
signature of boundary bound states in LDOS measurements we now consider
boundary conditions of the form $\Phi_\mathrm{c}(\tau,0)=0$,
$\Phi_\mathrm{s}(\tau,0)=\Phi_\mathrm{s}^0$ with
$0\le\Phi_\mathrm{s}^0\le\pi$. In terms of the right- and left-moving fields
this corresponds to $R_\sigma(\tau,0)=-e^{-\ii f_\sigma\Phi_\mathrm{s}^0/2}\,
L_\sigma(\tau,0)$, where $f_\up=1=-f_\dw$. We note that these boundary
conditions break spin rotational symmetry. If we go over to the case of the 1D
Mott insulator by exchanging spin and charge degrees of freedom the spin
rotational symmetry remains intact and the boundary conditions correspond to a
local potential. A discussion of the general case with nontrivial boundary
phase shifts in both spin and charge sectors is left for a longer publication.

\begin{figure}[b]
\centering
\includegraphics[scale=0.34,clip=true]{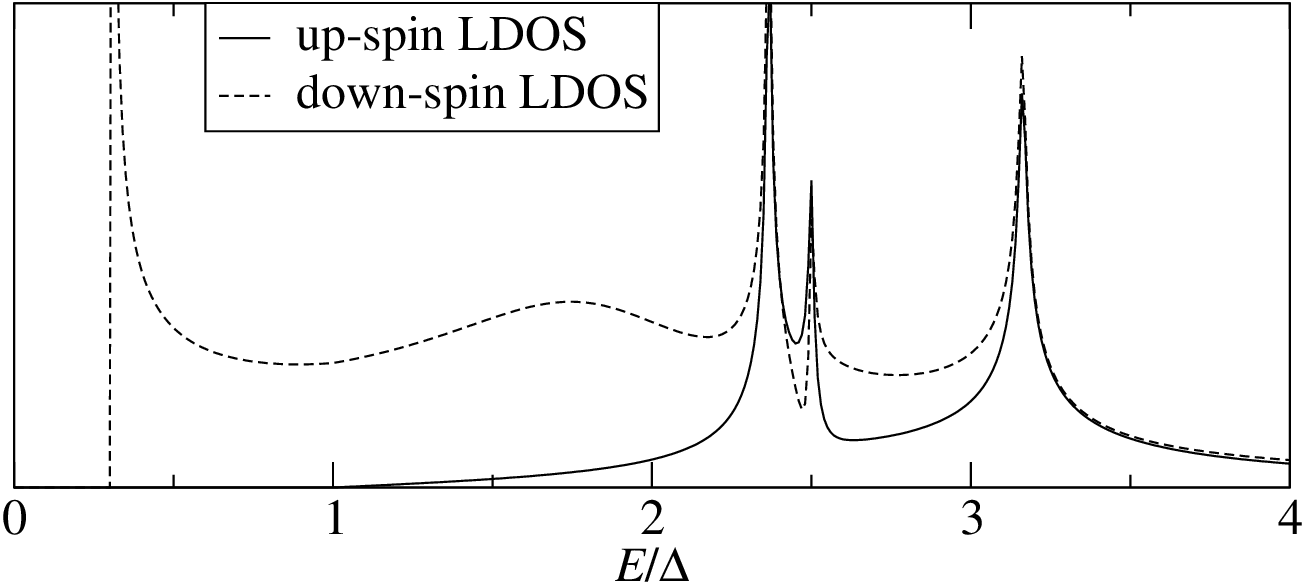}
\caption{$|N_\up(E,2k_\mathrm{F}+q)|$ (full line)
  and $|N_\dw(E,2k_\mathrm{F}+q)|$ (dashed line) for $v_\mathrm{s}q/\Delta=6$,
  $K_\mathrm{c}=1$, $K_\mathrm{s}=1/\sqrt{2}$, $v_\mathrm{s}=2v_\mathrm{c}$,
  and $\Phi_\mathrm{s}^0=0.9\,\pi$.  The singularity of
  $N_\dw(E,2k_\mathrm{F}+q)$ at $E=\Delta \sin\Phi_\mathrm{s}^0$ is due to the
  boundary bound state. The broad maximum of $N_\dw$ at $E\approx 1.8$ is due
  to the excitation of the boundary bound state and additional charge
  excitations.}
\label{fig:plot3}
\end{figure}

The leading terms in the form-factor expansion for the chiral Green's function
$G_{\sigma\sigma'}^{RL}$ are still given by \eqref{eq:GRLspin}, but now the
boundary reflection amplitude depends on $\sigma$. For example, at the
Luther-Emery point we have
$K^\sigma(\theta)=\sin(\ii\theta/2-f_\sigma\Phi_\mathrm{s}^0/2)/
\cos(\ii\theta/2+f_\sigma\Phi_\mathrm{s}^0/2)$.  We note that nevertheless the
Green's function is still diagonal in spin indices
$G_{\sigma\sigma'}^{RL}\propto\delta_{\sigma\sigma'}$.  If
$\pi/2\le\Phi_\mathrm{s}^0$, $K^\dw(\theta)$ has a pole in the physical strip
$0<\mathfrak{Im}\,\theta<\pi/2$, which gives rise to an additional term linear
in the boundary reflection matrix in the form-factor expansion. The resulting
contribution to $N_\dw(E,Q)$ has a non-dispersing singularity at the lower
threshold $\Delta \sin\Phi_\mathrm{s}^0$, see Fig.~\ref{fig:plot3}.  The
emergence of a non-dispersing feature within the gap signals the presence of a
boundary bound state. We note that the boundary bound state appears only in
the down-spin channel $N_\dw(E,Q)$. On the other hand, if we were to consider
$N_\sigma(E<0,Q)$, the additional feature would appear in the up-spin channel
only.

Finally we wish to discuss the implications of our results for STM
experiments on quasi-1D materials. Our results apply for energies
above the 1D-3D cross-over scale, which is set by the strength of the
3D couplings.
Perhaps the most interesting materials to which our findings may
be applied are two-leg ladders like
Sr$_{14}$Cu$_{24}$O$_{41}$~\cite{Abbamonte-04}. The model we have studied
captures the essential features of the low-energy description of (weakly
doped) two-leg ladders, namely a gapless charge sector and a gapped spin
sector. The tunneling current measured in STM experiments is directly related
to the local density of states $N_\sigma(E,Q)$. From the
$2k_\mathrm{F}$-component of the current one can hence extract informations
about the CDW correlations in the presence of an impurity.  In particular, we
expect the various peaks in $N_\sigma(E,Q)$ corresponding to the pinned CDW
order and the dispersing spin and charge degrees of freedom to appear in the
Fourier transform of the local differential conductance.  A possible asymmetry
$N_\up(E,Q)-N_\dw(E,Q)$ may be detected using a magnetic STM
tip~\cite{magnSTM}.

In summary, we have determined the low energy LDOS in strongly correlated
gapped 1D systems such as Mott insulators and CDW states in the presence of a
strong impurity potential. We have shown that the spatial Fourier transform of
the LDOS can be used to infer characteristic properties of the bulk state of
matter. The LDOS is dominated by a singularity at $2k_\mathrm{F}$, which is
indicative of the pinning of the CDW order at the position of the impurity.
The LDOS further exhibits clear signatures of propagating collective spin and
charge modes, which reflect the nature of the underlying electron-electron
interactions. We have investigated the modification of the LDOS in the
presence of impurity bound states and discussed the relevance of our results
to STM measurements on two-leg ladder materials like
Sr$_{14}$Cu$_{24}$O$_{41}$.

This work was supported by the DANL under grant BMBF-LPD 9901/8-145 (DS), the
EPSRC under grant EP/D050952/1 (FHLE), the NSF under grant DMR 0442537
(EF), the US DOE, Division of Basic Energy Sciences under Award
DE-FG02-07ER46453 (EF), the Frederick Seitz Materials Research
Laboratory at UIUC (EF, AJ) and the ESF network INSTANS.


\begin{thebibliography}{10}
\bibitem{endnote}
While the precise description of the spin sector for two-leg
ladders is considerably more complicated, we expect our calculation to
correctly capture the qualitative features.
\bibitem{endnote3}
We note that by virtue of the gap in the
spin sector a weak impurity potential renormalizes to strong coupling
even for moderate attractive interactions.
\end{thebibliography}
\end{document}